\documentclass{iau307}
\usepackage{graphicx}
\usepackage{natbib}
\usepackage{url}
\usepackage{dtklogos}
\bibpunct{(}{)}{;}{a}{}{,}

\title[Testing evolutionary models with binary stars] 
{Probing high-mass stellar evolutionary models with binary stars}

\author[A. Tkachenko]   
{A.\ Tkachenko$^1$
  \thanks{Postdoctoral Fellow of the Fund for Scientific Research (FWO), Flanders, Belgium}}

\affiliation{$^1$Instituut voor Sterrenkunde, KU Leuven,
Celestijnenlaan 200D, B-3001 Leuven, Belgium \\ email: {\tt
Andrew.Tkachenko@ster.kuleuven.be}}

\pubyear{2014}
\volume{307}
\pagerange{}
\jname{New windows on massive stars: asteroseismology, interferometry, and spectropolarimetry}
\editors{G. Meynet, C. Georgy, J.H. Groh \& Ph. Stee, eds.}

\begin{document}

\maketitle

\begin{abstract}
Mass discrepancy is one of the problems that is pending a solution
in (massive) binary star research field. The problem is often solved
by introducing an additional near core mixing into evolutionary
models, which brings theoretical masses of individual stellar
components into an agreement with the dynamical ones. In the present
study, we perform a detailed analysis of two massive binary systems,
V380\,Cyg and $\sigma$~Sco, to provide an independent, asteroseismic
measurement of the overshoot parameter, and to test state-of-the-art
stellar evolution models.

\end{abstract}

\firstsection 
\section{Introduction}

One of the major problems that is currently pending a solution in
binary star research field is the so-called mass discrepancy
problem. It stands for the difference between the component masses
inferred from binary dynamics (hereafter, dynamical masses) and
those obtained from spectral characteristics of stars and
evolutionary models (hereafter, theoretical masses). The mass
discrepancy problem observed in massive O- and B-type stars has been
known for more than 20 years already and has been discussed in
detail by \citet{Herrero1992}. \citet{Hilditch2004} pointed out that
the discrepancy does not disappear when the effects of rotation are
included into the models.

A remarkable mass discrepancy has been reported by
\citet{Guinan2000} for the primary components of the V380\,Cyg
system. The authors showed that large amount of core overshoot
($\alpha_{ov}$ = 0.6 pressure scale height) can account for the
difference between dynamical and theoretical mass of the primary
component. This large amount of overshoot contradicts the typical
value of $\alpha_{ov} <= 0.2$~Hp observed for single stars of
similar mass (see e.g. Aerts 2013, Aerts~et~al. 2003, 2011; Briquet
et al. 2011). Moreover, the second largest value after V380\,Cyg of
$\alpha_{ov}\sim$0.45 has also been measured in a binary system, for
the 8~M$_{\odot}$ primary component of the $\theta$~Ophiuchi system
\citep{Briquet2007}. Recently, \citet{Garcia2014} found that
convective overshoot $\alpha_{ov}>$0.35 is required to reproduce
observed absolute dimensions of both components of the V578\,Mon
system. Thus, there seems to be a tendency of measuring larger core
overshoot in binary stars than in single objects, within the same
stellar mass range. The reason is that this parameter often
effectively accounts for the above mentioned mass discrepancy, but
we need to investigate the feasibility of using core overshoot alone
to explain such a complex problem as discrepancy between dynamical
and theoretical masses in binary stars.

In this paper, we present a study of two massive binary star
systems, V380\,Cyg and $\sigma$~Sco, and aim at measuring core
overshoot parameter for pulsating components using asteroseismic
methods.

\begin{table}
\begin{center}
\caption{Key orbital, physical, and atmospheric parameters of the
V380\,Cyg and $\sigma$~Sco systems, as derived from our photometric
and/or spectroscopic data.} \label{tab1}
\begin{tabular}{ll|cccc|}\hline
\multicolumn{2}{c|}{\textbf{Parameter}} & \multicolumn{2}{c}{\textbf{V380\,Cyg}} & \multicolumn{2}{c|}{\textbf{$\sigma$~Sco}}\\
& & \textbf{Primary} & \textbf{Secondary} & \textbf{Primary} & \textbf{Secondary}\\
\hline
Period, & (day) & \multicolumn{2}{c}{12.425719} & \multicolumn{2}{c|}{33.016$\pm$0.012}\\
Periastron passage time, & (HJD) & \multicolumn{2}{c}{24\,54\,602.888$\pm$0.007} & \multicolumn{2}{c|}{24\,34\,886.11$\pm$0.04}\\
Periastron long., & (degree) & \multicolumn{2}{c}{138.4$\pm$0.4} & \multicolumn{2}{c|}{288.1$\pm$0.8}\\
eccentricity, &  & \multicolumn{2}{c}{0.2261$\pm$0.0004} & \multicolumn{2}{c|}{0.383$\pm$0.008}\\
RV semi-amplitude, & (km\,s$^{-1}$) & 93.54$\pm$0.07 & 152.71$\pm$0.22 & 30.14$\pm$0.35 & 47.01$\pm$0.98\\
Mass, & (M$_{\odot}$) & 11.43$\pm$0.19 & 7.00$\pm$0.14 & 14.7$\pm$4.5 & 9.5$\pm$2.9\\
Radius, & (R$_{\odot}$) & 15.71$\pm$0.13 & 3.819$\pm$0.048 & 9.2$\pm$1.9 & 4.2$\pm$1.0\\
$T_{\rm eff}$, & (K) & 21\,700$\pm$300 & 22\,700$\pm$1\,200 & 25\,200$\pm$1\,500 & 25\,000$\pm$2\,400\\
$\log$\,$g$, & (dex) & 3.104$\pm$0.006 & 4.120$\pm$0.011 & 3.68$\pm$0.15 & 4.16$\pm$0.15\\
v\,$\sin$\,i, & (km\,s$^{-1}$) & 98$\pm$2 & 38$\pm$2 & 31.5$\pm$4.5 & 43.0$\pm$4.5\\
\hline
\end{tabular}
\end{center}
\end{table}

\section{V380\,Cyg\label{V380Cyg}}

V380\,Cyg is a bright ($V$ = 5.68) double-lined spectroscopic binary
\citep[SB2,][]{Hill1984} consisting of two B-type stars residing in
an eccentric 12.4 day orbit. The primary component is an evolved
star, whereas the secondary just started its life on the
main-sequence. \citet{Pavlovski2009} revisited the {\it U, B, V}
light curves obtained by \citet{Guinan2000} and collected about 150
high-resolution \'{e}chelle spectra using several telescopes. The
authors presented a revised orbital solution, and used spectral
disentangling technique \citep{Simon1994} formulated in Fourier
space \citep{Hadrava1995}, as implemented in the {\sc fdbinary} code
\citep{Ilijic2004}, to determine disentangled spectra of both binary
components. Similar to the results of \citet{Guinan2000}, a
remarkable mass discrepancy was found for the primary component of
V380\,Cyg. Moreover, \citet{Pavlovski2009} came to the same
conclusion as \citet{Hilditch2004} did, namely that the effects of
rotation included into evolutionary models could not fully account
for the observed discrepancy.

The discovery of seismic signal in the primary component of
V380\,Cyg \citep{Tkachenko2012} opened up an opportunity of
obtaining an independent measurement of the overshoot parameter for
this binary system. We base our analysis on about 560 days long
time-series of high-precision {\it Kepler} photometry, and about 400
high-resolution, high signal-to-noise ratio (S/N) spectra obtained
with {\sc hermes} spectrograph \citep{Raskin2011} attached to
1.2-meter Mercator telescope (La Palma, Canary Islands).

The effects of binarity in the {\it Kepler} light curve were
modelled using the {\sc jktwd} wrapper \citep{Southworth2011} of the
2004 version of the Wilson-Devinney code
\citep{Wilson1971,Wilson1979}. The time-series of spectra were
analysed with the {\sc fdbinary} code to determine spectroscopic
orbital solution and disentangled spectra of both stellar
components. We found that light curve yields poor constraints on the
shape of the orbit, because of strong correlation between
eccentricity $e$ and longitude of periastron $\omega$. Since the
quantities $e\,\cos \omega$ and $e\,\sin \omega$ are respectively
well constrained from photometry and spectroscopy, we constrained
the orbital shape by iterating between the two analyses: the light
curve was used to determine best fit $\omega$ for a given $e$, and
the spectral disentangling to find the best $e$ for a given
$\omega$. Analysis of the disentangled spectra delivered accurate
atmospheric parameters and individual abundances for both binary
components. Table~\ref{tab1} lists some key orbital, physical, and
atmospheric properties of this binary system; for more details,
reader is referred to \citet{Tkachenko2014a}.

\begin{figure*}[t]
\includegraphics[scale=0.8]{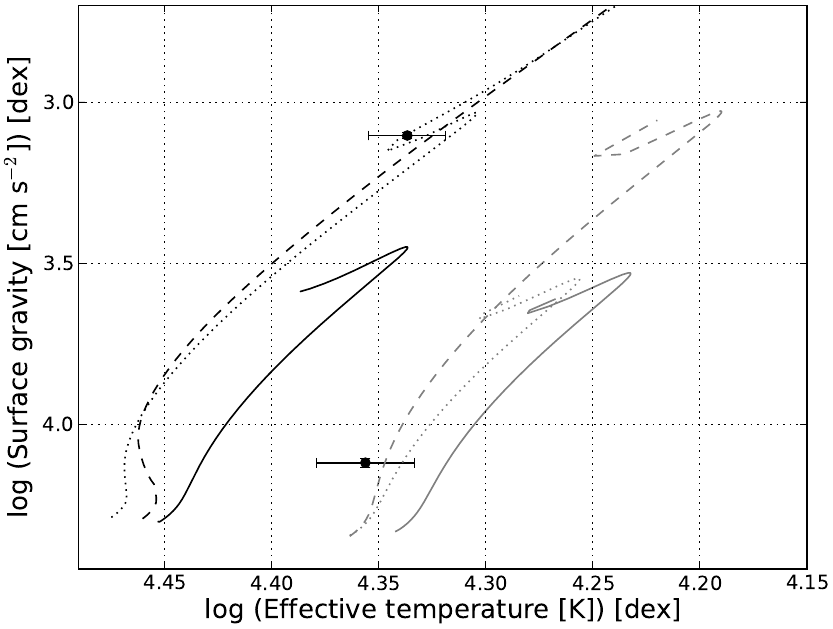}
\includegraphics[scale=0.8]{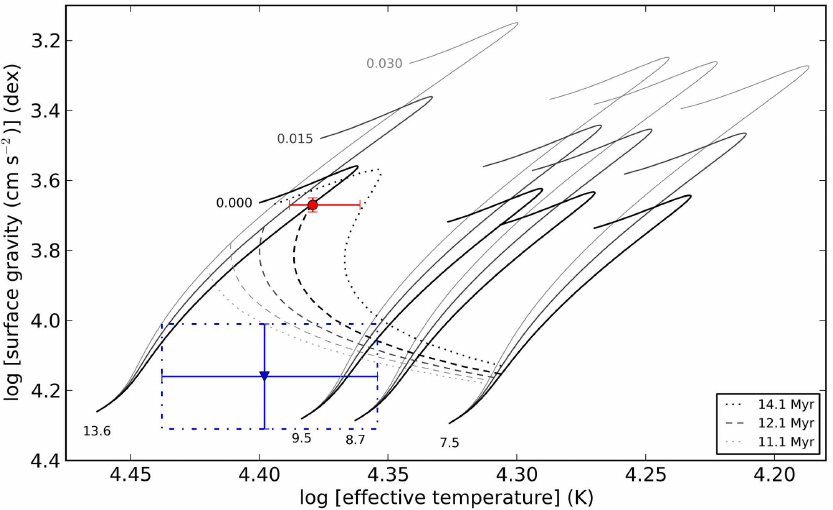}
\caption{\textbf{Left:} Location of the primary and secondary
components of V380\,Cyg in the $T_{\rm eff}$–-$\log$\,$g$ diagram.
Solid, long- and short-dashed tracks correspond to models 1, 2 and 3
in Table~\ref{tab2}, respectively. $T_{\rm eff}$ and $\log$\,$g$
values are those from Table~\ref{tab1}. \textbf{Right:} Position of
the primary (circle) and the secondary (triangle) of the
$\sigma$~Sco system in the $T_{\rm eff}$–-$\log$\,$g$ diagram, along
with the {\sc mesa} evolutionary tracks. The initial masses as well
as the overshoot parameter $f_{ov}$ are indicated in the plot. The
atmospheric parameters $T_{\rm eff}$ and $\log$\,$g$ are those from
Table~\ref{tab3}. The isochrones corresponding to the age of the
system of 12.1~Myr and its error bars deduced from seismology of the
primary are indicated by the dashed lines.} \label{fig1}
\end{figure*}

\begin{table}[t]
\begin{center}
\caption{Evolutionary model parameters for both components of the
V380\,Cyg system. $\alpha_{ov}$ and $\upsilon$ stand for the
overshoot parameter and initial rotation rate, respectively.}
\label{tab2}
\begin{tabular}{ll|llllll|}\hline
\multicolumn{2}{c|}{\textbf{Parameter}} & \multicolumn{3}{c}{\textbf{Primary}} & \multicolumn{3}{c|}{\textbf{Secondary}}\\
& & Model 1 & Model 2 & Model 3 & Model 1 & Model 2 & Model 3\\
\hline
M, & (M$_{\odot}$) & 11.43 & 12.00 & 12.90 & 7.00 & 7.42 & 7.42\\
$Z$, & (dex) & 0.014 & 0.012 & 0.012 & 0.014 & 0.012 & 0.012\\
$\alpha_{ov}$, & (H$_{\rm p}$) & 0.2 & 0.6 & 0.3 & 0.2 & 0.6 & 0.2\\
$\upsilon$, & (km\,s$^{-1}$) & 0 & 241 & 243 & 0 & 0 & 0\\
Age, & (Myr) & -- & 21.5 & 18 & -- & 21.5 & 18\\
\hline
\end{tabular}
\end{center}
\end{table}

Photometric residuals obtained after the subtraction of our best fit
model were subjected for frequency analysis. The majority of the
frequencies are variable both in appearance and amplitude, in
agreement with the conclusions made by \citet{Tkachenko2012} about
stochastic nature of the signal. The variability consistent with the
expected rotation period of the primary component has been detected
on top of the binarity and stochastic oscillation signals in the
star. We speculate that this signal comes from rotational modulation
of spot-like chemical abundance or temperature structures on the
surface of the primary component. To verify this hypothesis,
spectral line profiles of the primary has been examined for spot
signatures, after subtracting the contribution of the companion star
from the composite spectra of V380\,Cyg. We found a remarkable
variability in all observed silicon lines of more massive binary
component, which could not be detected in spectral lines of other
chemical elements and was found to be consistent with the rotational
period of the star. {\sc invers8} \citep{Piskunov1993} code was used
to perform Doppler Imaging analysis based on several prominent lines
of doubly ionized silicon found in the spectrum of the primary
component. The obtained results suggest the presence of two
high-contrast stellar surface abundance spots which are located
either at the same latitude or longitude.

Finally, we compare our revised fundamental stellar parameters of
both components of the V380\,Cyg system with the state-of-the-art
evolutionary models computed with the {\sc mesa} code
\citep{Paxton2011,Paxton2013}. Figure~\ref{fig1} (left) shows the
location of both components of V380\,Cyg in the $T_{\rm
eff}$–-$\log$\,$g$ diagram along with the evolutionary tracks. The
two models that fit the positions of both stars in the diagram,
taking into account the error bars, are illustrated by long- and
short-dashed lines (models 2 and 3 in Table~\ref{tab2},
respectively). The dynamical mass models for both binary components
are shown by solid lines (model 1 in Table~\ref{tab2}). The {\sc
mesa} model predictions clearly point to mass discrepancy for the
primary component, in agreement with the findings by
\citet{Guinan2000} and \citet{Pavlovski2009}. \textbf{We conclude
that present-day single-star evolutionary models are inadequate for
this particular binary system, and lack a serious amount of
near-core mixing.}

\section{$\sigma$~Sco}

Sigma Scorpii is a double-lined spectroscopic binary in a quadruple
system. Two components are early B-type stars, residing in an
eccentric, 33 day period orbit. Though the star was a subject of
numerous photometric and spectroscopic studies in the middle of 20th
century, its double-lined nature was discovered by
\citet{Mathias1991}. So far, the studies by \citet{Mathias1991},
\citet{Pigulski1992}, and \citet{North2007} have been the most
extensive ones focusing on orbital and physical properties of the
system.

Besides the $\sigma$~Sco system is a spectroscopic binary, its
evolved primary component is known to be unstable to
$\beta$~Cep-type stellar pulsations. Moreover, according to
\citet{Kubiak1980}, the amplitude of the dominant, radial pulsation
mode of the primary is comparable to the orbital semi-amplitude
$K_1$ of this star. This fact was not taken into account in either
of the previous studies focusing on orbital solution, but is taken
into consideration in our study.

Our analysis is based on some 1000 high-resolution spectra collected
with the CORALIE spectrograph attached to the 1.2-meter Euler Swiss
Telescope (La Silla, Chile). Orbital parameters of the system were
initially derived based on an iterative approach, and further on
refined using the method of spectral disentangling. The spectral
disentangling was applied to a couple of dozen carefully selected
spectra and corresponding to a zero pulsation phase (unperturbed
profile), since the method assumes no variability intrinsic to
stellar components forming a binary system. For more details on the
procedure, reader is referred to \citet{Tkachenko2014b}. The
disentangled spectra were used to determine accurate atmospheric
parameters and chemical composition of both stars. The final set of
orbital parameters as well as the spectroscopically derived values
of $T_{\rm eff}$ and $\log$\,$g$ are given in Table~\ref{tab1}. The
masses and radii listed in this table were determined from our
orbital parameters and interferometric value of the orbital
inclination angle reported by \citet{North2007}.

\begin{table}[t]
\begin{center}
\caption{Fundamental parameters of both components of the
$\sigma$~Sco, after seismic modelling of the primary. Parameters
determined spectroscopically are highlighted in boldface.}
\label{tab3}
\begin{tabular}{ll|ll|}\hline
\multicolumn{2}{c|}{\textbf{Parameter}} & \multicolumn{1}{c}{\textbf{Primary}} & \multicolumn{1}{c|}{\textbf{Secondary}}\\
\hline
Mass, & (M$_{\odot}$) & 13.5$^{+0.5}_{-1.4}$ & 8.7$^{+0.6}_{-1.2}$\\
Radius, & (R$_{\odot}$) & 8.95$^{+0.43}_{-0.66}$ & 3.90$^{+0.58}_{-0.36}$\\
Luminosity ($\log$\,($L$)), & (L$_{\odot}$) & 4.38$^{+0.07}_{-0.15}$ & 3.73$^{+0.13}_{-0.15}$\\
Age of the system, & (Myr) & \multicolumn{2}{c|}{12.1}\\
Overshoot ($f_{ov}$), & (H$_{\rm p}$) & 0.000$^{+0.015}$ & \multicolumn{1}{c|}{--}\\
$T_{\rm eff}$, & (K) & 23\,945$^{+500}_{-990}$ & \textbf{25\,000$^{+2\,400}_{-2\,400}$}\\
$\log$\,$g$, & (dex) & 3.67$^{+0.01}_{-0.03}$ & \textbf{4.16$^{+0.15}_{-0.15}$}\\
\hline
\end{tabular}
\end{center}
\end{table}

We further made use of the fact that the primary component of
$\sigma$~Sco is a radial mode pulsator, and performed asteroseismic
analysis of this star. Evolutionary models were computed with the
{\sc mesa} code, while p- and g-mode eigenfrequencies for mode
degrees $l=0$ to 3 have been calculated in the adiabatic
approximation with the {\sc gyre} stellar oscillation code
\citep{Townsend2013}. The addition of the seismic constraints to the
modelling implied a drastic reduction in the uncertainties of the
fundamental parameters, and provided an age estimate (see
Table~\ref{tab3}). Figure~\ref{fig1} (right) shows the position of
both components of the $\sigma$~Sco system in the $T_{\rm
eff}$--$\log$\,$g$ diagram, along with the evolutionary tracks. The
error bars are those obtained from the evolutionary models and
3$\sigma$ spectroscopic uncertainties for the primary and secondary,
respectively. \textbf{Though we make an a priori assumption in our
seismic modelling that the models are appropriate for the primary
component, similar to the case of V380\,Cyg, we still find mass
discrepancy for the main-sequence secondary component of the
$\sigma$~Sco system.}

\bibliographystyle{iau307}
\bibliography{MyBiblio}

\begin{thebibliography}{}

\bibitem[\protect\astroncite{{Aerts}}{2013}]{Aerts2013}
{Aerts}, C. 2013,
\newblock {\em EAS Publications Series} 64, 323

\bibitem[\protect\astroncite{{Aerts et al.}}{2003}]{Aerts2003}
{Aerts}, C., {Thoul}, A., {Daszy{\'n}ska}, J., et al. 2003,
\newblock {\em Science} 300, 1926

\bibitem[\protect\astroncite{{Aerts et al.}}{2011}]{Aerts2011}
{Aerts}, C., {Briquet}, M., {Degroote}, P., et al. 2011,
\newblock {\em \aap} 534, A98

\bibitem[\protect\astroncite{{Briquet et al.}}{2007}]{Briquet2007}
{Briquet}, M., {Morel}, T., {Thoul}, A., et al. 2007,
\newblock {\em \mnras} 381, 1482

\bibitem[\protect\astroncite{{Briquet et al.}}{2011}]{Briquet2011}
{Briquet}, M., {Aerts}, C., {Baglin}, A., et al. 2011,
\newblock {\em \aap} 527, A112

\bibitem[\protect\astroncite{{Garcia et al.}}{2014}]{Garcia2014}
{Garcia}, E.~V., {Stassun}, K.~G., {Pavlovski}, K., et al. 2014,
\newblock {\em \aj} 148, 39

\bibitem[\protect\astroncite{{Guinan et al.}}{2000}]{Guinan2000}
{Guinan}, E.~F., {Ribas}, I., {Fitzpatrick}, E.~L., et al. 2000,
\newblock {\em \apj} 544, 409

\bibitem[\protect\astroncite{{Hadrava}}{1995}]{Hadrava1995}
{Hadrava}, P. 1995,
\newblock {\em \aaps} 114, 393

\bibitem[\protect\astroncite{{Herrero et al.}}{1992}]{Herrero1992}
{Herrero}, A., {Kudritzki}, R.~P., {Vilchez}, J.~M., et al. 1992,
\newblock {\em \aap} 261, 209

\bibitem[\protect\astroncite{{Hilditch}}{2004}]{Hilditch2004}
{Hilditch}, R.~W. 2004,
\newblock {\em ASP Conference Series} 318, 198

\bibitem[\protect\astroncite{{Hill \& Batten}}{1984}]{Hill1984}
{Hill}, G. \& {Batten}, A.~H. 2004,
\newblock {\em \aap} 141, 39

\bibitem[\protect\astroncite{{Iliji\'{c} et al.}}{2004}]{Ilijic2004}
{Iliji\'{c}}, S., {Hensberge}, H., {Pavlovski}, K., \& {Freyhammer}, L.~M. 2004,
\newblock {\em ASP Conference Series} 318, 111

\bibitem[\protect\astroncite{{Kubiak}}{1980}]{Kubiak1980}
{Kubiak}, M. 1980,
\newblock {\em \actaa} 30, 41

\bibitem[\protect\astroncite{{Mathias et al.}}{1991}]{Mathias1991}
{Mathias}, P., {Gillet}, D., \& {Crowe}, R. 1991,
\newblock {\em \aap} 252, 245

\bibitem[\protect\astroncite{{North et al.}}{2007}]{North2007}
{North}, J.~R., {Davis}, J., {Tuthill}, P.~G., et al. 2007,
\newblock {\em \mnras} 380, 1276

\bibitem[\protect\astroncite{{Pavlovski et al.}}{2009}]{Pavlovski2009}
{Pavlovski}, K., {Tamajo}, E., {Koubsk{\'y}}, P., et al. 2009,
\newblock {\em \mnras} 400, 791

\bibitem[\protect\astroncite{{Paxton et al.}}{2011}]{Paxton2011}
{Paxton}, B., {Bildsten}, L., {Dotter}, A., et al. 2011,
\newblock {\em \apjs} 192, 3

\bibitem[\protect\astroncite{{Paxton et al.}}{2013}]{Paxton2013}
{Paxton}, B., {Cantiello}, M., {Arras}, P., et al. 2013,
\newblock {\em \apjs} 208, 4

\bibitem[\protect\astroncite{{Pigulski}}{1992}]{Pigulski1992}
{Pigulski}, A. 1992,
\newblock {\em \aap} 261, 203

\bibitem[\protect\astroncite{{Piskunov \& Rise}}{1993}]{Piskunov1993}
{Piskunov}, N.~E. \& {Rice}, J.~B. 1993,
\newblock {\em \pasp} 105, 1415

\bibitem[\protect\astroncite{{Raskin et al.}}{2011}]{Raskin2011}
{Raskin}, G., {van Winckel}, H., {Hensberge}, H., et al. 2011,
\newblock {\em \aap} 526, A69

\bibitem[\protect\astroncite{{Simon \& Sturm}}{1994}]{Simon1994}
{Simon}, K.~P. \& {Sturm}, E. 2009,
\newblock {\em \aap} 281, 286

\bibitem[\protect\astroncite{{Southworth et al.}}{2011}]{Southworth2011}
{Southworth}, J., {Zima}, W., {Aerts}, C., et al. 2011,
\newblock {\em \mnras} 414, 2413

\bibitem[\protect\astroncite{{Tkachenko et al.}}{2012}]{Tkachenko2012}
{Tkachenko}, A., {Aerts}, C., {Pavlovski}, K., et al. 2012,
\newblock {\em \mnras} 424, L21

\bibitem[\protect\astroncite{{Tkachenko et al.}}{2014a}]{Tkachenko2014a}
{Tkachenko}, A., {Degroote}, P., {Aerts}, C., et al. 2014a,
\newblock {\em \mnras} 438, 3093

\bibitem[\protect\astroncite{{Tkachenko et al.}}{2014b}]{Tkachenko2014b}
{Tkachenko}, A., {Aerts}, C., {Pavlovski}, K., et al. 2014b,
\newblock {\em \mnras} 442, 616

\bibitem[\protect\astroncite{{Townsend \& Teitler}}{2013}]{Townsend2013}
{Townsend}, R.~H.~D. \& {Teitler}, S.~A. 2013,
\newblock {\em \mnras} 435, 3406

\bibitem[\protect\astroncite{{Wilson}}{1979}]{Wilson1979}
{Wilson}, R.~E. 1979,
\newblock {\em \apj} 234, 1054

\bibitem[\protect\astroncite{{Wilson \& Devinney}}{1971}]{Wilson1971}
{Wilson}, R.~E. \& {Devinney}, E.~J. 1971,
\newblock {\em \apj} 166, 605

\end{thebibliography}

\end{document}